# Voting Data-Driven Regression Learning for Discovery of Functional Materials and Applications to Two-Dimensional Ferroelectric Materials

**Xing-Yu Ma[1], Hou-Yi Lyu[1, 2], Xue-Juan Dong[1], Zhen Zhang[1], Kuan-Rong Hao[1], Qing-Bo Yan[2*], Gang Su[3, 1, 4*]**

[1]School of Physical Sciences, University of Chinese Academy of Sciences, Beijing 100049, China.

[2]Center of Materials Science and Optoelectronics Engineering, College of Materials Science and Optoelectronic Technology, University of Chinese Academy of Sciences, Beijing 100049, China.

[3]Kavli Institute for Theoretical Sciences, and CAS Center of Excellence in Topological Quantum Computation, University of Chinese Academy of Sciences, Beijing 100190, China.

[4]Physical Science Laboratory, Huairou National Comprehensive Science Center, Beijing 101400, China.

**Regression machine learning is widely applied to predict various materials. However, insufficient materials data usually leads to a poor performance. Here, we develop a new voting data-driven method that could generally improve the performance of regression learning model for accurately predicting properties of materials. We apply it to investigate a large family (2135) of two-dimensional hexagonal binary compounds focusing on ferroelectric properties and find that the performance of the model for electric polarization is indeed greatly improved, where 38 stable ferroelectrics with out-of-plane polarization including 31 metals and 7 semiconductors are screened out. By an unsupervised learning, actionable information such as how the number and orbital radius of valence electrons, ionic polarizability, and electronegativity of constituent atoms affect polarization was extracted. Our voting data-driven method not only reduces the size of materials data for constructing a reliable learning model but also enables to make precise predictions for targeted functional materials.**

## Introduction

Machine learning has been widely applied to materials and chemical sciences,[1,2,3] such as designing perovskites,[4,5] predicting properties of inorganic crystals[6] and catalytic properties of alloyed nanoclusters,[7] discovery of organic photovoltaic molecules,[8] two-dimensional (2D) optoelectronic materials,[9] and 2D ferromagnetic materials, [10] *etc*. However, the advances of machine learning for materials discovery are suffered from some constraints. For instance, high-precision machine learning model requires a sufficient amount of data, but the limited variety of materials may lead to inadequate data, and huge resource consumption also suppresses massive collection of materials data. More importantly, the functional materials data is usually limited or even insufficient, hindering a wide application of machine learning for the discovery of advanced functional materials. Recently, it is proposed that the data-driven methodology may relieve above restriction and has spurred several applications for predictions of battery cycle life[11] and chemical reactivity, [12] 2D ferromagnetic materials, [10] 2D ferroelectric materials, [13] and electrocatalysts. [14] Up to now, this technique is mainly used to improve the accuracy of classification machine learning model, which depends on an iteration criterion that prioritizes the decision-making process on next unexplored data that are iteratively added into training dataset for retraining.[10-13, 15] Unlike classification learning that identifies candidates belong to a certain group, regression machine learning can predict the properties of each material, and the latter is more important for functional material discovery than the former. However, data-driven methodology has not been applied to regression learning due to the lack of proper iteration criteria, thus hindering further applications of regression learning for predicting and designing functional materials.

In this work, we propose a new voting data-driven regression learning approach, which could generally improve the performance of the regression learning model for accurately predicting better properties of functional materials. The method includes a voting process of five different regression learning algorithms that select next new data. The successful application of this method on the 2D hexagonal binary compounds (HBCs) demonstrates the performance of regression learning model for electric polarization can be remarkably improved under the restriction of limited materials data. With the above high-precision model, the electric polarization of the whole family (2135) of 2D HBCs has been studied, where 38 stable ferroelectric materials including 31 metals and 7 semiconductors are screened out. Besides, we classify the 2135 materials into two classes with high and low polarization by utilizing unsupervised machine learning and extract an actionable principle component for polarization that clarifies how the number and orbital radius of valence electrons, ionic polarizability, and electronegativity of the constituent atoms affect the polarization. The successful application of our method for the discovery of 2D ferroelectric HBCs indicates that it could generally achieve a high-precision regression learning model based on a relatively small amount of materials data for designing a wide range of functional materials.

## Results and Discussion

### Voting data-driven regression learning method

A great challenge for the application of machine learning in materials science is that inadequate materials data always lead to poor performance of the model. The data-driven scheme employs an iterative strategy that may be taken as a reliable choice, which adds new data into the training dataset to improve the performance of machine learning model in next iteration. The principle is that these new data have the greatest impact on the performance of model due to different contributions of each data to the model.[15] The key of the method is the iteration criteria, which determine how these new data are chosen from the total dataset. For example, in a classification learning model, the iteration criterion is defined as the prediction probability obtained by the classification learning model such that new data with predicted probability in a certain range are added into the training dataset.[10-13] Regression learning model can make predictions for the targeted materials properties, which is very critical for the design of functional materials, however, current regression learning models do not have such proper iteration criteria as they cannot give prediction probabilities. Here, we provide a scheme to address this challenge by introducing the voting data-driven scheme and a new iteration criterion for regression learning model. Specially, five different regression learning algorithms are adopted to make predictions, respectively, and the next new data are chosen by a voting process of these algorithms, i.e., the candidates with the largest prediction error are selected as next new data adding to the training dataset for retraining. These added dataset plays the most critical role in improving the performance of model than other dataset. This can be viewed as five regression learning expert votes to determine which data are the next most important data for retraining. Notably, the voting process does not involve specific materials and properties but only depends on the voting results of five different regression learning algorithms. Therefore, this method could be general and suitable for various properties of many functional materials. The detailed procedure is described as follows.

Fig. 1 shows the voting data-driven regression learning method. A fraction of the total dataset is randomly selected as initial dataset for training and test dataset, which contains the proposed features (descriptors) and targeted properties of materials generated from density functional theory (DFT) calculations.[16] Five different regression learning algorithms widely used in materials science, i.e., gradient boosting regression (GBR), [17] adaboost regression (ABR), [18] bagging, [19] random forest regression (RFR) [20] and light gradient boosting machine (LGBM) [21] algorithms are adopted to train five different models, respectively, which establish different mappings between features and targeted properties. For the remaining dataset, the above five models are used to predict the targeted properties and to evaluate the standard error for each material. The standard error has the form of $\sigma = \sqrt{[(y_{GBR} - \bar{y})^2 + (y_{ABR} - \bar{y})^2 + (y_{Bag} - \bar{y})^2 + (y_{RFR} - \bar{y})^2 + (y_{LGBM} - \bar{y})^2]/5}$ , where $y_{GBR}$, $y_{ABR}$, $y_{Bag}$, $y_{RFR}$ and $y_{LGBM}$ are predicted values of the above five models, respectively, and $\bar{y}$ is the average value of them. During the voting process, the top 30 materials with the maximum standard error are selected and added to the initial training/test dataset for retraining. With the updated training/test dataset, new models were then obtained with improved precision. The above process is repeated until the coefficient of determination value ($R^2$) converges, which occurs at multiple iterations. Finally, the model with the best $R^2$ and mean absolute error (MAE) are chosen as the optimal model, with which the targeted properties of remaining unexplored materials are predicted for further investigation. More importantly, the above process only relies on the standard error, which depends on the above five different algorithms, and it is irrelevant to special materials and targeted properties. Thus, the scheme proposed here could be widely applicable to improve the performance of regression learning model for various properties of a broad range of functional materials.

## Application to discovery of novel 2D ferroelectric materials

Recently, 2D ferroelectric materials [22-25] have begun to attract great attention due to their promising applications in ferroelectric nonvolatile memory, [26] ferroelectric field effect transistor[27] and ferroelectric photocatalysts.[28] Inspired by the experimentally synthesized single-layer hexagonal $GdAg_2$, $GdAu_2$ and $MgB_2$, [29-31] here we focus on the family of 2D HBCs, which could be ferroelectric due to the potential spontaneous inversion symmetry breaking. For instance, hexagonal $CrB_2$ has been predicted to possess both ferroelectricity and ferromagnetism.[32] In order to evaluate the capability and validity of above approach, we apply the method to investigate the 2D HBCs family and discover the ferroelectricity among them. Fig. 2(a) illustrates the schematic structure of 2D HBCs ($MX_2$, M and X are different atoms), which can be considered as M atoms (red balls) embedded in hexagonal lattice of X atoms (yellow balls). As shown in Fig. 2(c), if M atoms are in the hexagonal plane, the structure has space group *P6/mmm* (No.191) (*Cmmm* (No.65) for some materials), which corresponds to non-polar point groups *6/mmm* (or *mmm*) and we denote it as "high-symmetric phase". There are possible spontaneous inversion symmetry breaking as shown in Fig. 2(b) or 2(d), where M atoms are above or below the hexagonal plane and $d$ is the vertical distance between M and the hexagonal plane of X atoms. Meanwhile, the space group reduces to *P6mm* (No. 183) (or *Cmm2* (No. 35)), which corresponds to polar point groups *6mm* (or *mm2*), and we denote it as "low-symmetric phase". High-symmetric phase and low-symmetric phase correspond to paraelectric phase and ferroelectric phase, respectively. Based on the above 2D HBCs prototype, we generate diverse 2D HBCs in our practice by replacing M atoms and X atoms by 50 elements across the periodic table (see Table S1). Note that $MX_2$ and $XM_2$ structures are stoichiometrically inequivalent and there are total 2450 $MX_2$ structures. Data cleaning is a method to ensure the consistency of data by removing abnormal and unnecessary data before machine learning training. By performing a full relaxed geometric structure optimization, we discovered that two X atoms of 315 structures in 2450 $MX_2$ are not sitting on the same plane, resulting in that they do not possess a hexagonal lattice (see Fig. S1), and thus we only considered the remaining 2135 $MX_2$ structures as a total dataset to ensure the consistency of data.

213 of 2D HBCs are randomly selected as the initial training/test dataset for training initial regression learning model. These materials are described with 22 initial features (listed in Table S2), and their optimized geometric structures and electric polarization are calculated by using Vienna *ab initio* simulation package (VASP) [33] (the details of calculations can be found in supplementary information). With the voting data-driven method described above, an iterative scheme is introduced to obtain the high-precision model. Fig. 3(a) shows the performance of model at each iteration for five different algorithms. Obviously, $R^2$ obviously increases and MAE decreases with the increase of iterations, revealing that the voting data-driven scheme indeed improves the performance distinctly. At the 15th iteration, the $R^2$ of GBR model converges at the highest value (0.872) with the lowest MAE (0.040) (Figs. 3(a)), which corresponds to the optimal model comparing with others regression models, indicating that to enhance the accuracy of the regression model is reliable.

The way to reduce the complexity of machine learning model is to choose a suitable number of features that could perfectly reflect the properties of materials. We utilized the GBR algorithm to perform a feature reduction engineering, in which the features with less impact are successively excluded (see Fig. S4) and found that 16 features are sufficient to construct the optimal feature space with the highest $R^2$ (0.883) and the lowest MAE (0.038) of the model (Fig. 3(b)) of the model. The optimal features mainly include the displacements (*d*) and 15 element-related properties of constituent atoms, i.e., orbital radii of the inner valence electron, $r_{M(in)}$ and $r_{X(in)}$; orbital radii of the outer valence electron, $r_{M(out)}$ and $r_{X(out)}$; the electronegativities (Martynov-Batsanov scale[34]), $E_M$ and $E_X$; ionization energies, $IE_M$ and $IE_X$; the number of valence electron, $n_M$ and $n_X$; ionic radii, $r_M$ and $r_X$; ionic polarizability of M, $P_M$; the number of inner valence electron of M, $n_{M(in)}$, and structural factor features to distinguish two different $MX_2$ and $XM_2$ structures, $S_f = \frac{r_M + r_X}{2r_X}$, where $r_M$ and $r_X$ are ion radii for M and X, respectively (see Table S2 for details). Then we acquire the importance of features by GBR algorithm, and the results reveal that *d* and $S_f$ play the most critical roles in determining the polarization, followed by $r_{X(in)}$ and $r_{X(out)}$ (Fig. 3(c)). Fig. 3(d) shows Pearson correlation coefficient matrices, which indicate the positive and negative correlations between feature pairs. Interestingly, $IE_M$ has highly negative correlation with $r_{M(out)}$, which is consistent with the case that larger $IE_M$ implies the decreasing ability to lose outer valence electrons, resulting in smaller $r_{M(out)}$.

To evaluate the validity of our new method, we applied the obtained optimal model to predict the electric polarization of the remaining 1472 materials. As illustrated in Fig. 3 (e), the comparison between the voting data-driven regression learning-predicted and DFT-calculated results reveal that the $R^2$ and MAE are 0.854 and 0.029, respectively, which are consistent well with that of the optimal model on the training/test dataset. Figure 3 (f) shows that the absolute errors of more than 80% of the materials are within $\pm$0.05 eÅ/ unit cell. These results prove that the validity of the above method is very reasonable. We also compared the performance of our method with GBR (without using the voting data-driven method) and found that the $R^2$ of our method is larger than the latter (Table S3), suggesting that our method could indeed improve the performance. Apparently, we may conclude that the voting data-driven regression learning method provides a reliable scheme to improve the performance of regression model and a feasible way to accelerate searching for advanced functional materials with a reasonable accuracy.

**Actionable information extraction**

Unlike supervised machine learning that makes accurate predictions for the targeted materials properties or labels through training on dataset with existing properties or labels, unsupervised machine learning could draw boundaries between different clusters and extract information from the features regardless of whether the properties or labels exist or not, such as agglomerative hierarchical clustering algorithm (AHC) [35] and principal component analysis (PCA)[36], etc. AHC can identify the candidates similar to the cluster with high ionic conductivity through training on all materials containing lithium ion, [37] and PCA could extract actionable information for phase transition[38] and magnetoelectric effect.[39] We utilized AHC method to train a bottom-up clustering model of 2135 2D HBCs with only features based on element-related properties of the constituent atoms. It turns out that this large family can be classified into three clusters, as shown in Fig. 4 (a). Specially, the clustering shows a good quality as the materials in the same group have similar characteristics, and different clusters are well distinguished (Fig. S9(l)). In Fig. 4(b), the violin plot of polarization for cluster I shows obviously high polarizations while that of clusters II and III have similar shapes and low polarizations. We reduced the three clusters into two groups, i.e., the high polarization cluster I and low polarization clusters II and III. To check the robustness of clustering, we also adopted K-means clustering algorithm[40] and Gaussian mixture model clustering algorithm [41] to cluster 2135 2D HBCs with the same features as AHC. Although the principles of the three clustering algorithms are different, the results are almost consistent with each other (Fig. S7), indicating the strong robustness of the unsupervised learning clustering of high polarization and low polarization materials using only the features based on element-related properties.

The key of PCA is to find a new set of feature vectors that capture the most essential information by linearly combining initial features with weights, i.e., the 'principal components', and from which the actionable information can be extracted. To obtain concise and essential 'principal components', we need to reduce the number of inputting initial features for PCA. Since the unsupervised learning algorithm such as PCA cannot give out the importance of features, we train a new regression learning model to obtain the importance with only the features based on element-related properties by GBR algorithm (Figs. S8 and S9). After a feature reduction engineering aiming at the target of best clustering quality, we obtained a new optimal feature space only containing 13 most important features (see supplemental information for details of the features). Then we utilized the PCA technique to capture the essential information and extract two principle components, which are constructed by linear combinations of weighted contributions of the 13 features based on element-related properties, thereby significantly reducing the dimensionality of the dataset from thirteen to two. Fig. 4(c) shows the reduced two-dimensional feature space, in which a distinct boundary between cluster I and cluster II/III materials can be observed. Besides, the boundary line is nearly vertical to Component1, implying that Component1 has much greater influence on polarization than Component2. The polarization versus the Component1 for 2135 2D HBCs materials (see Fig. S10) also reveals that the polarizations roughly increase with the increase of Component1. Component1 can be expressed as $0.785\mathrm{P_M} - 0.126\mathrm{P_X} + 0.002\mathrm{E_X} - 0.01\mathrm{E_M} + 0.530\mathrm{r_{M(out)}} - 0.084\mathrm{r_{X(out)}} - 0.017\mathrm{n_M} + 0.003\mathrm{n_X}$, revealing that if $\mathrm{P_X}$, $\mathrm{E_M}$, $\mathrm{n_M}$, and $\mathrm{r_{X(out)}}$ decrease, and $\mathrm{P_M}$, $\mathrm{E_X}$, $\mathrm{n_X}$, and $\mathrm{r_{M(out)}}$ increase, then Component1 and polarization will increase. In short, Component1 shows how these element-related properties of the constituent atoms affect the polarization, which overcomes the black box nature of supervised learning models. The sure independence screening and sparsifying operator (SISSO) is a supervised learning algorithm that can also extract critical information based on the compressed sensing technique.[42] We have also tried the SISSO algorithm and found that PCA algorithm shows better performance here in the classification of materials with high and low polarizations (Fig. S11). The unsupervised learning method adopted here provides a reliable way to extract the actionable or chemical information for designing other functional materials.

## Ferroelectrics

As shown in the bottom panel of Fig. S2, we screened out 38 stable ferroelectrics with the criteria of polarization above 0.015 eÅ /unit cell and verified the dynamical stabilities using the finite displacement method using a 5×5×1 supercell.[43] To the end, we investigated systematically the electronic structures of all 38 ferroelectric HBCs by using the hybrid functional based on the Hyed-Scuseria-Ernzerhof exchange-correlation functional, [44] finding 31 metals and 7 semiconductors (more properties can be found in Table S4). We further calculated the energy versus polarization profiles under different perpendicular electric fields for $HgPt_2$ and $YAu_2$ (Figs. S20), and the results show 1.0 and 0.8 V/Å could be the critical reversible electric fields, respectively, which are nearly the same order of magnitude as that of 2D $In_2Se_3$ (0.66 V/Å), [22] suggesting their electric polarizations could be reversed by a proper vertical external electric field.

To understand why some ferroelectrics $MX_2$ composed by two types of metal atoms are semiconductors, we employed the SISSO algorithm and obtained a new feature for identifying ferroelectrics as metals or semiconductors (Fig. S13 (a)). The feature has the form $\mathrm{Feature1} = (n_{X(in)} \times IE_M)/(n_M + n_X)$, where $n_{X(in)}$ and $n_X$ are the number of inner valence electrons and valence electrons of X atoms, respectively, and $IE_M$ and $n_M$ are ionization energy and the number of valence electrons for M atoms, respectively. More importantly, we discovered that the probability to become a ferroelectric semiconductor increases when $\mathrm{Feature1}$ increases (Fig. S13 (b)). If $n_M$ and $n_X$ of a ferroelectric material decrease while $IE_M$ increases, the number of total valence electrons of the material and the ability to lose electrons of M atoms will decrease, resulting in the decrease of the number of itinerant electrons that shows more ionic bonding character between M and X atoms, and thereby they could be more likely to be semiconductors.

## Ferroelectric metals

In 1965, Anderson and Blount firstly proposed the concept of ‘ferroelectric metal’, pointing out that the electric polarization may appear in certain martensitic transitions due to the inversion symmetry breaking, in contrast to the conventional belief that ferroelectricity and metallicity could not coexist in a metal because conducting electrons of metals always screen out the internal electric dipole moment.[45] Recently, the coexistence of ferroelectricity and metallicity have been observed experimentally in two or three layers $WTe_2$, [46] and 2D $CrB_2$, CrN,[32] and some 2D bimetal phosphates ferroelectric metals [13] were also predicted. The above 31 ferroelectric metals can be divided into two groups (Group M and X) according to the observation that the conducting electrons densities $\rho_c(\vec{r})$ (partial electron densities within energy range $|E- E_f|< 0.05$ eV, and the details can be found in supplementary information) mainly distribute around M or X atoms (Table S4). Here, we focus on $YAl_2$ and $CaRh_2$ as typical examples of Group M and X, respectively. In Figs. 5 (a) and (b), the energy bands of $YAl_2$ and $CaRh_2$ are crossing the Fermi level, indicating that they are metals. In Figs. 5 (c) and (d), the projected electronic density of states (PDOS) reveal that the electronic states at the Fermi level are mainly contributed by the *d*-orbital electrons of Y atoms and Rh atoms, respectively. As shown in Figs. 5 (e) and (f), $\rho_c(\vec{r})$ of $YAl_2$ and $CaRh_2$ exhibits a typical *d*-orbital character of Y atoms and Rh atoms, respectively, which are both accordance with the observations from their PDOS. To visualize the conducting electron density on x-y plane, we define a ‘reduced’ conducting electron density $\rho_c(x,y) = \int \rho_c(\vec{r})dz$ i.e., integrating $\rho_c(\vec{r})$ over the *z* direction. As illustrated in Figs. 5(g) and (h), the ‘reduced’ conducting electron density (the blue surfaces) reveals clearly that the conducting electrons are mainly distributed around Y atoms and Rh atoms, respectively. Besides, to describe the spatial distribution of electrons that contribute to the electric polarization, a ‘reduced’ charge density difference $\rho_p(x,y) = \int [\rho_{FE}(\vec{r}) - \rho_{PE}(\vec{r})]dz$ was also introduced, where $\rho_{FE}(\vec{r})$ and $\rho_{PE}(\vec{r})$ are the total electron densities of a ferroelectric material in ferroelectric and paraelectric phases, respectively. Thus, we term $\rho_p(x,y)$ as the reduced 'FE-PE electron density difference' because it can clearly reflect the net electron density related to the electronic contribution to polarization. As shown by the red surfaces in Figs. 5(g) and (h), $\rho_p(x,y)$ exhibits an oscillating behavior and displays the spatial distribution of polarization in overall *x*-*y* plane. It is critical to note that for both materials $\rho_c(x,y)$ deviates obviously from that of $\rho_p(x,y)$ in *x*-*y* plane, i.e., the conducting electrons and 'FE-PE electrons density difference' are spatially separated in *x*-*y* plane (Figs. 5 (g) (h)), clearly manifesting the underlying mechanism of the coexistence of ferroelectricity and metallicity in these 2D metals, which is different from the spatial separation of 2D bimetal phosphates that occurs in the *z* direction.[13]

To understand why the conducting electrons are mainly distributed either around M atoms (Group M) or X atoms (Group X), we adopted the SISSO algorithm and acquired a new feature to identify which ferroelectric metals belong to Group M or Group X (Fig. S17), say, $\mathrm{Feature2} = |\mathrm{r_{X(in)}} - \mathrm{r_X}|/(\mathrm{E_M} - \mathrm{E_X})$, where $\mathrm{r_{X(in)}}$ and $\mathrm{r_X}$ are the orbital radius of inner valence electrons and ionic radius of X atoms, respectively, $\mathrm{E_M}$ and $\mathrm{E_X}$ are the electronegativities of M and N, respectively. Interestingly, when $\mathrm{E_M}$ is smaller than $\mathrm{E_X}$, the ability to attract electrons of M atoms is weaker than that of X atoms, where electrons are mainly distributed around X atoms, resulting that 'FE-PE electron density difference' is also mainly predominant around X atoms. Thus, the conducting electrons primarily distribute around M atoms according to the spatial separation and the ferroelectric metal will belong to Group M, otherwise, they belong to Group X. The conducting electrons mainly distribute around the atoms that usually have weaker electronegativity, while the low-density areas appear around another atoms (Figs. 5 (e) and (f)), implying that the conducting electrons do not completely exclude the external electric fields, which thus rationalize the coexistence of ferroelectricity and metallicity.

## Conclusion

In summary, we presented a new voting data-driven regression learning method, which only depends on the vote of different regression learning algorithms and could be generally used to enhance the performance of regression learning model for discovery of a broad range of advanced functional materials. With this method, we showed that the performance of the regression learning model for acquiring the electric polarizations of 2D HBCs can be indeed dramatically improved with only limited training dataset. Based on the high-precision regression learning model, we screened out 38 stable ferroelectric materials including 31 ferroelectric metals and 7 ferroelectric semiconductors from 2135 HBCs. In addition, by using the unsupervised machine learning, we succeeded in clustering the materials with high and low polarizations and obtained an actionable principle component for polarization, which elaborates how the number and orbital radius of valence electrons, ionic polarizability, and electronegativity of constituent atoms affect practically the electric polarization, a key character of ferroelectric materials. We applied the SISSO algorithm to obtain the novel interpretable descriptors for exposing the underlying mechanism of the origin and characteristics of metallicity, which contain the number of valence electrons and electronegativity of constituent atoms. Our approach provides an efficient and accurate but yet successful strategy that may greatly alleviate the shortage of the available materials data nowadays, which could dramatically accelerate to develop a wide of advanced functional materials.

## Methods

The density functional theory first-principles calculations are performed by projected augmented wave (PAW)[47] implemented Vienna ab initio simulation package (VASP).[33] The exchange-correlation interactions are treated using Perdew-Burke-Ernzerhof generalized gradient approximation (PBE-GGA).[48] The cut-off energy of 450 eV was set for the plane-wave basis and $10 \times 10 \times 1$ k-points are used to sample the Brillouin zone. The convergence criteria were $1\times10^{-6}$ eV for the energy difference in the electronic self-consistent calculation and $1\times10^{-3}$ eV/Å for the residual forces on lattice geometries and atomic positions. All electronic structures are calculated by using the hybrid functional based on the Hyed-Scuseria-Ernzerhof exchange-correlation functional.[44] Meanwhile, the on-site Coulomb correlated potential for 3d, 4d and 5d transition metals were chosen as U = 4 eV, 2.5 eV and 0.5 eV, respectively, which are reasonable for them,[49, 50] thus, the electronic structures calculations were also performed by PBE-GGA+U.

## Acknowledgements

This work is supported in part by the National Key R&D Program of China (Grant No. 2018YFA0305800), the Strategic Priority Research Program of CAS (Grant No. XDB28000000), the NSFC (Grant No. 11834014), Beijing Municipal Science and Technology Commission (Grant No. Z191100007219013), and University of Chinese Academy of Sciences. The calculations were performed on Era at the Supercomputing Center of Chinese Academy of Sciences and Tianhe-2 at National Supercomputing Center in Guangzhou.

## Author contributions

Qing-Bo Yan and Gang Su conceived the project and supervised the research. Xing-Yu Ma performed all calculations. Xing-Yu Ma, Qing-Bo Yan, and Gang Su developed the voting data-driven regression machine learning model, analyzed the results, and wrote the manuscript. Hou-Yi Lyu, Xue-Juan Dong, Zhen Zhang, and Kuan-Rong Hao participated in the discussion.

## Competing interests

The authors declare that they have no conflict of interest.

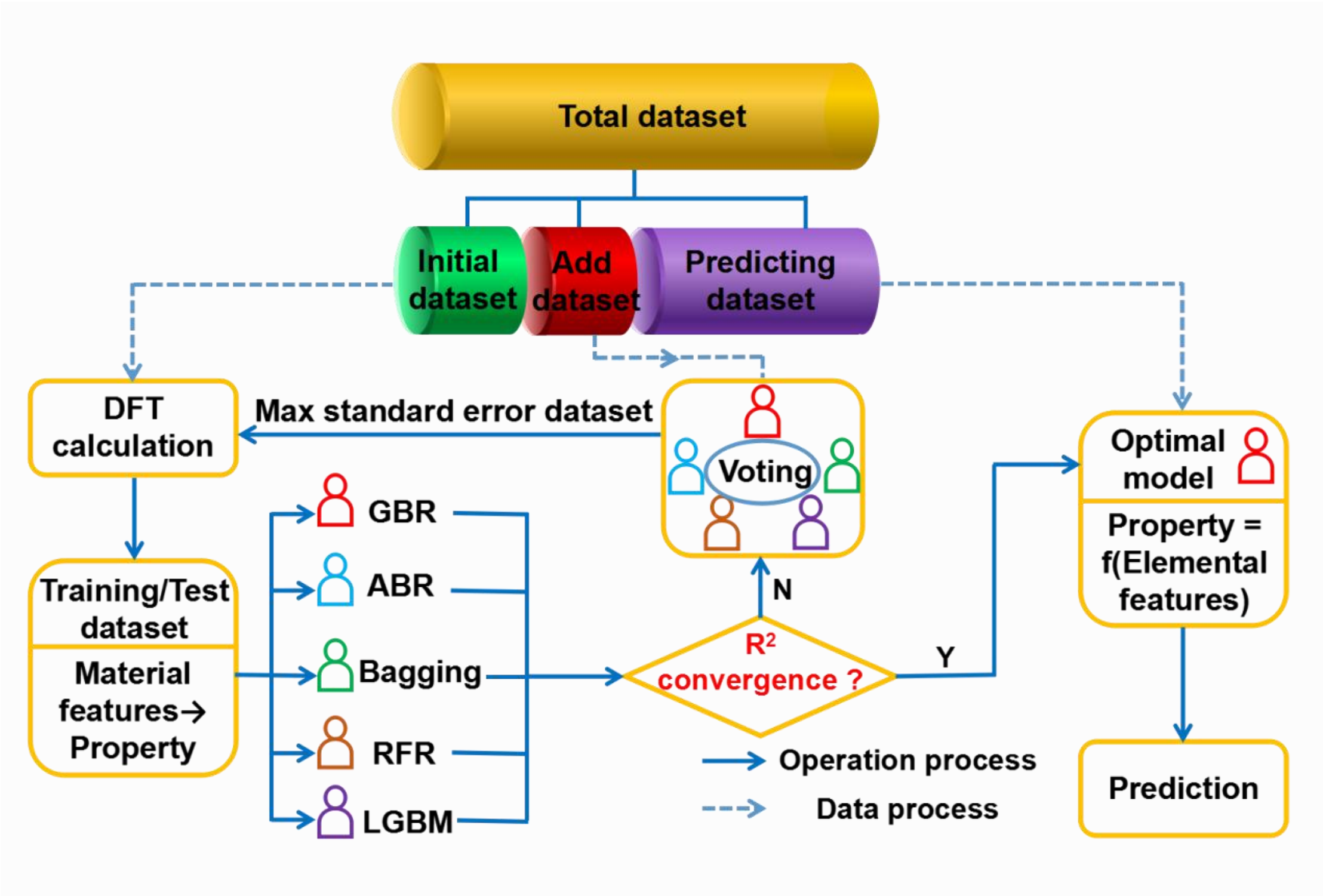

Figure 1. The training and prediction projects of voting data-driven regression learning method. The cylinders represent different datasets. Red, cyan, green, brown and purple represent the gradient boosting regression (GBR), adaboost regression (ABR), bagging (Bagging), random forest regression (RFR) and light gradient boosting machine (LGBM) algorithms, respectively. “Max standard error dataset” represents the top 30 dataset with maximum standard errors, which are evaluated by the predicted values of the above five regression learning algorithms.

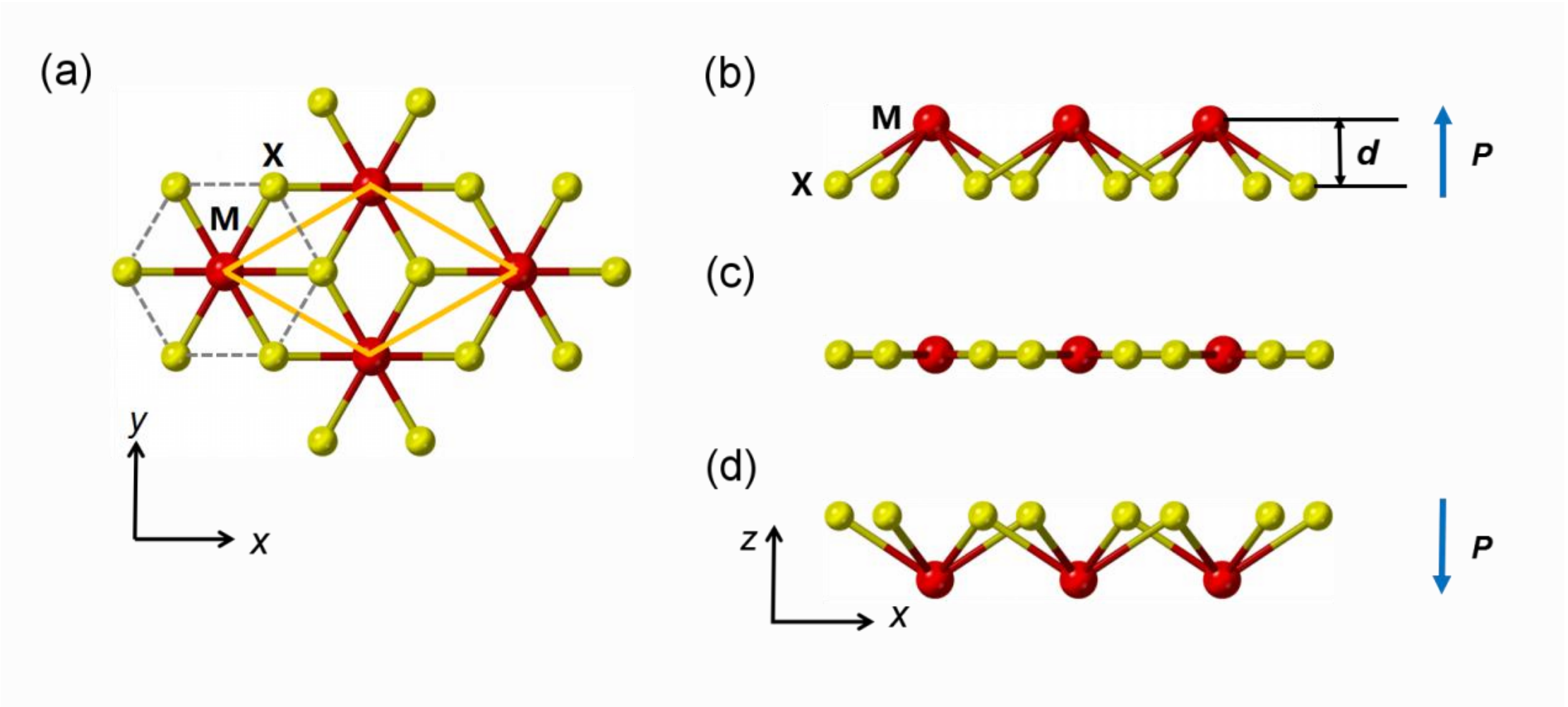

Figure 2. Schematic structure of 2D hexagonal binary compounds ($MX_2$, the M and X atoms are different atoms). (a) Top view. The yellow parallelogram indicates the unit cell and gray dash lines denote a honeycomb lattice formed by X atoms. (c) Side view of the high-symmetry phase. (b) and (d) Side views of low-symmetry phases with different polarization directions. $d$ is the vertical relative displacement between M atoms and X atoms. The blue arrows (up/down) indicate the directions of out-of-plane electric polarization (***P***). The red and yellow balls represent M atoms and X atoms, respectively.

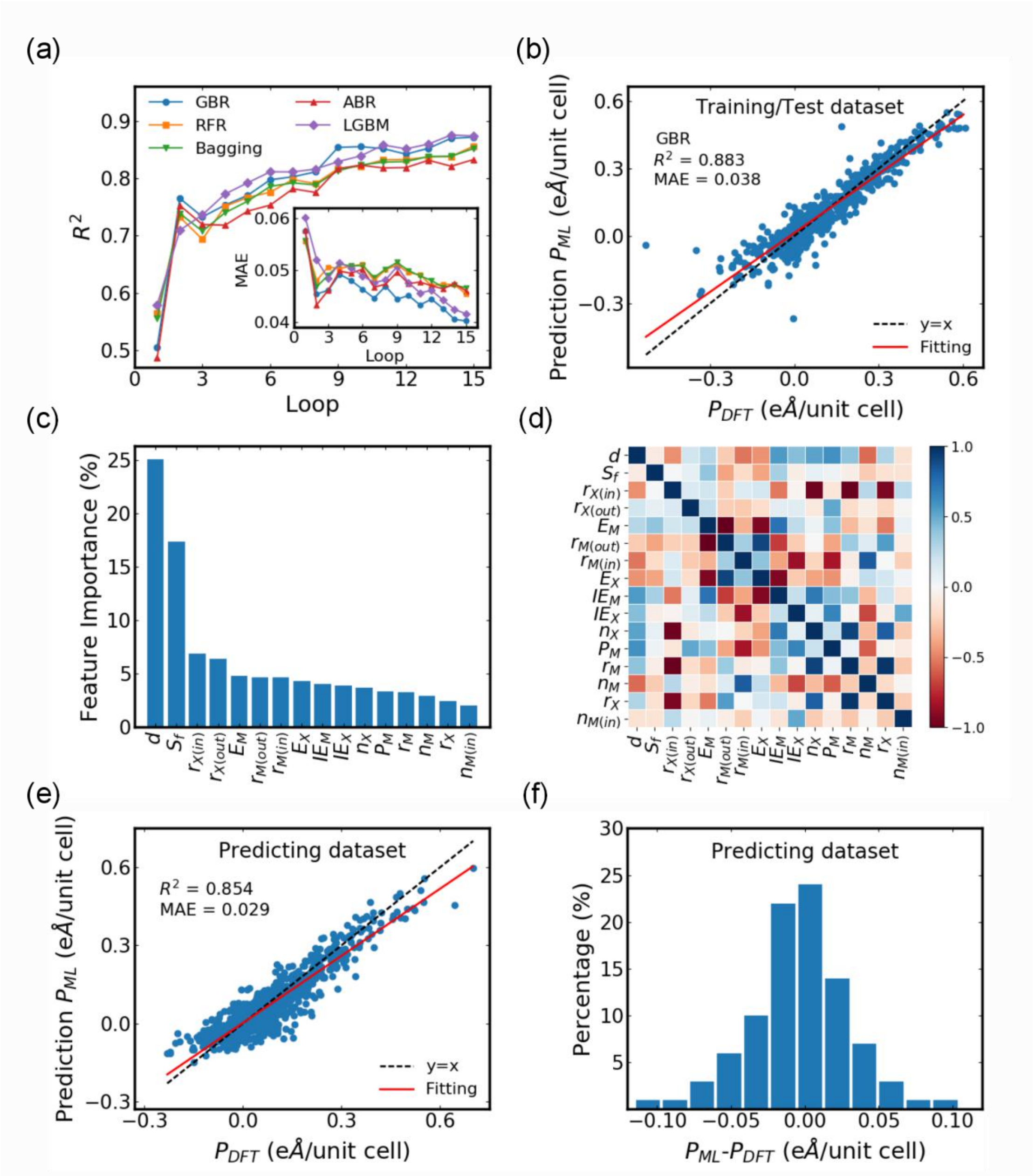

Figure 3. The results of voting data-driven regression learning model, importance and correlation of 16 optimal features. (a) The results of voting data-driven regression learning model. (b) The fitting results of the polarizations between DFT-calculated values and ten-fold cross-validation predicted values for the training/test dataset. (c) Features are ranked with importance using GBR algorithm. (d) Statistical heat map shows the correlation of these features. (e) The fitting results of the polarization between the DFT-calculated values and the voting data-driven regression learning model predicted ones for predicting dataset (1472 materials). The reliability of the model is evaluated by the coefficient of determination ($R^2$) and the mean absolute error (MAE). The red line is the fitting curve and the gray dash line is the guidance

line whose prediction error is zero. (f) Fraction of materials according to their errors between the voting data-driven regression learning-predicted and DFT-calculated polarizations.

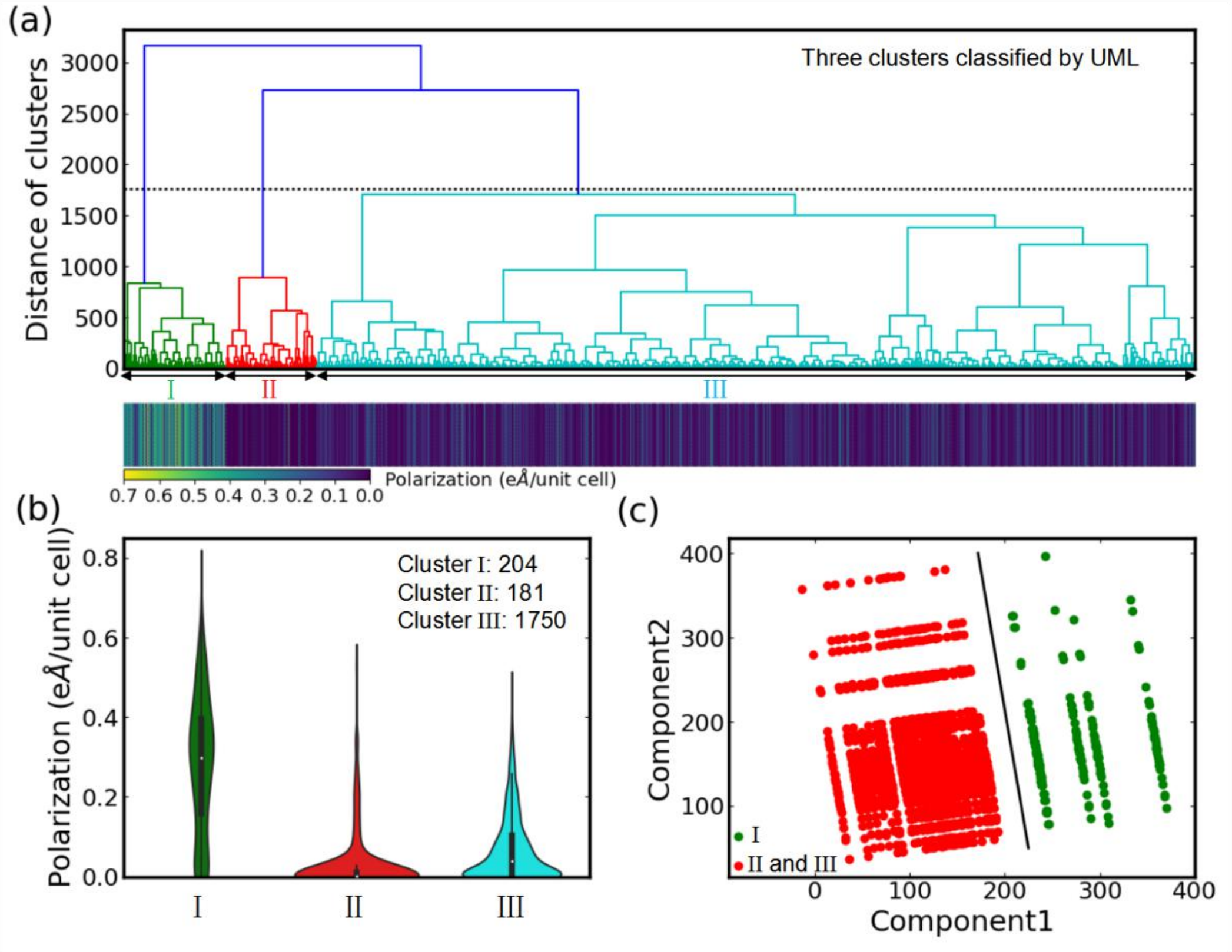

Figure 4. Unsupervised machine learning (UML) clustering of 2135 HCBs. (a) Bottom-up tree dendrogram of three clusters of 2135 HCBs obtained using the agglomerative hierarchical clustering algorithm. The dashed line shows the position where all materials are partitioned into three clusters, marked as I–III from left to right in different colors. The spectral line maps the dendrogram to the polarization, showing the clustering of 2135 HCBs. The color bar shows the scale of polarization. (b) Violin plots of polarization of three clusters. The outer shells of the violins bound all data, where the width represents the number of materials, narrow horizontal lines bound 95% of the data, thick horizontal lines bound 50% of the data, and white dots represent medians, respectively. (c) Visualization of all materials in the two principle components by PCA method (cluster I: green, cluster II/III: red). The black line represents the boundary of classification. Note that the two components are linear combinations of the weighted contributions of the element-related properties of constituent atoms.

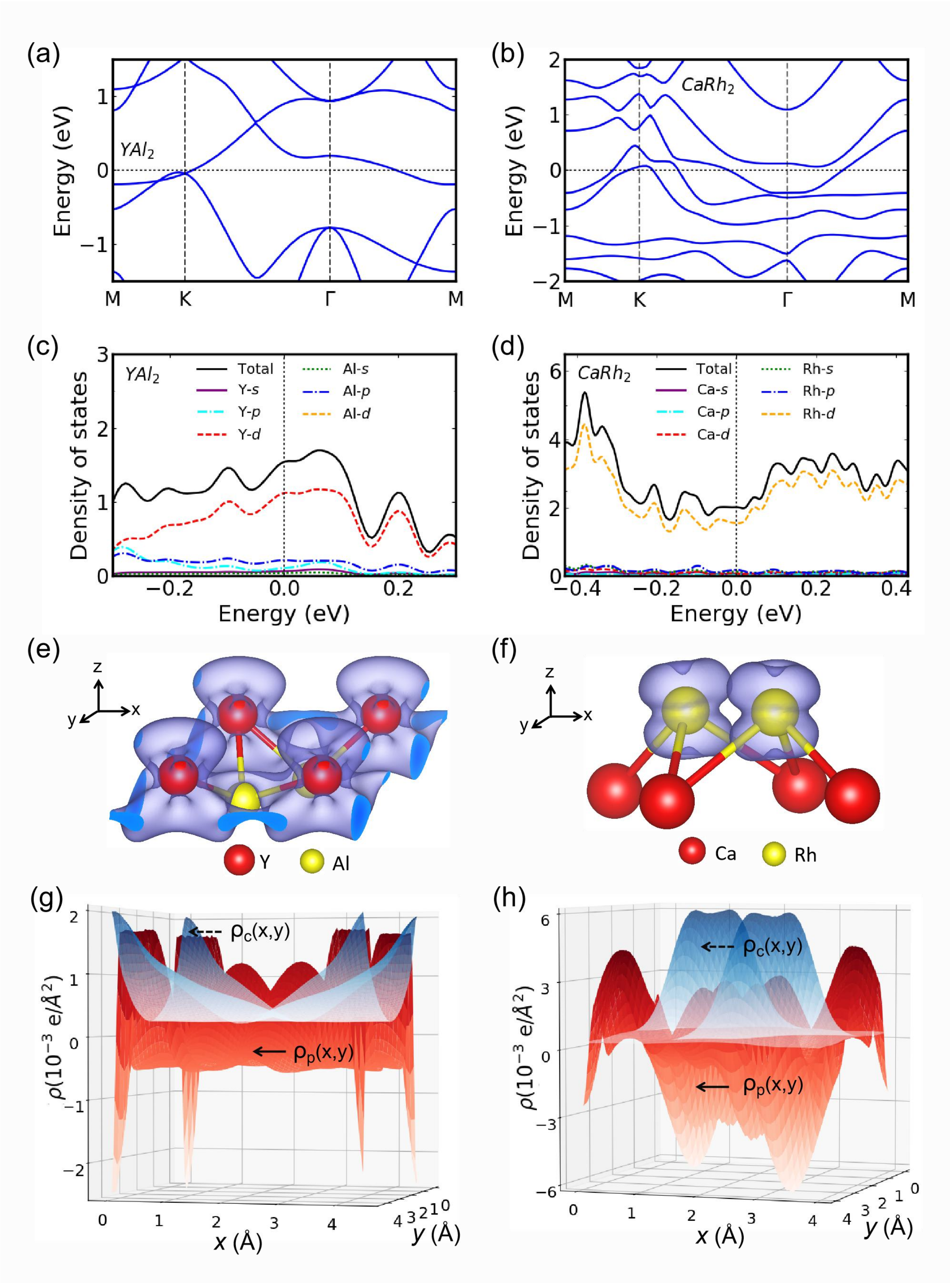

Figure 5. Electronic properties of two ferroelectric metals: $YAl_2$ and $CaRh_2$. (a) and (b) Band structures at PBE+U level for two materials. (c) and (d) Projected density of states (PDOS). (e) and (f) Schematic views of partial electron densities within the energy range $|E - E_f| < 0.05$ eV, i.e., $\rho_c(\vec{r})$, respectively. (g) and (h) Blue surface indicates the reduced partial electron density $\rho_c(x,y)$, and red surface indicates the reduced "FE-PE electron density difference" $\rho_p(x,y)$, where the $x$ and $y$ axis denote the spatial $x$ and $y$ coordinates the same as the geometrical structures in (e) and (f).